\documentclass[pra,preprint,doublecolumn,superscriptaddress, nobalancelastpage]{revtex4-1}

\usepackage{tikz}
\newcommand*\circled[1]{\tikz[baseline=(char.base)]{
            \node[shape=circle,draw,inner sep=1pt,fill=black] (char) {#1};}}

\usepackage{etoolbox}
\patchcmd{\section}
  {\centering}
  {\raggedright}
  {}
  {}
\patchcmd{\subsection}
  {\centering}
  {\raggedright}
  {}
  {}

\usepackage{physics}
\usepackage{tcolorbox}
\usepackage[sort&compress]{natbib}
\usepackage{soul}
\usepackage{graphicx}
\usepackage{epstopdf}
\usepackage{verbatim}
\usepackage{float}
\usepackage{sidecap}
\sidecaptionvpos{figure}{c}
\usepackage{lipsum}
\usepackage{setspace}
\usepackage{dcolumn}
\usepackage{bm}
\usepackage{amssymb}
\usepackage{amsmath}
\usepackage{siunitx}
\usepackage{mathtools}
\usepackage{color}
\usepackage{multirow}
\usepackage{lipsum}
\usepackage[a4paper]{geometry}
\newgeometry{top=1.25cm,right=1.5cm,left=2cm,bottom=1.25cm}
\usepackage[colorlinks=true, urlcolor=blue, pdfborder={0 0 0}]{hyperref}
\hypersetup{
linkcolor=blue,
citecolor=blue
}

\usepackage{graphicx}
\usepackage{bm}
\usepackage{hyperref}

\begin{document}


\title{Comparative analysis of error mitigation techniques for variational quantum eigensolver implementations on IBM quantum system}

\author{Shaobo Zhang}
\email{shaozhang@student.unimelb.edu.au}
\affiliation{School of Computing and Information Systems, The University of Melbourne, Parkville, 3010, Australia}
\affiliation{School of Physics, The University of Melbourne, Parkville, 3010, Australia}

\author{Charles D. Hill}
\email{cdhill@unimelb.edu.au}
\affiliation{School of Physics, The University of Melbourne, Parkville, 3010, Australia}
\affiliation{School of Mathematics and Statistics, The University of Melbourne, Parkville, 3010, Australia}

\author{Muhammad Usman}
\email{musman@unimelb.edu.au}
\affiliation{School of Computing and Information Systems, The University of Melbourne, Parkville, 3010, Australia}
\affiliation{School of Physics, The University of Melbourne, Parkville, 3010, Australia}
\affiliation{Data61, CSIRO, 3168, Clayton Australia}

\begin{abstract}

Quantum computers are anticipated to transcend classical supercomputers for computationally intensive tasks by exploiting the principles of quantum mechanics. However, the capabilities of the current generation of quantum devices are limited due to noise or errors, and therefore implementation of error mitigation and/or correction techniques is pivotal to reliably process quantum algorithms. In this work, we have performed a comparative analysis of the error mitigation capability of the [[4,2,2]] quantum error-detecting code (QEC method), duplicate circuit technique, and the Bayesian read-out error mitigation (BREM) approach in the context of proof-of-concept implementations of variational quantum eigensolver (VQE) algorithm for determining the ground state energy of H$_2$ molecule. Based on experiments on IBM quantum device, our results show that the duplicate circuit approach performs superior than the QEC method in the presence of the hardware noise. A significant impact of cross-talk noise was observed when multiple mappings of the duplicate circuit and the QEC method were implemented simultaneously -- again the duplicate circuit approach overall performed better than the QEC method. To gain further insights into the performance of the studied error mitigation techniques, we also performed quantum simulations on IBM system with varying strengths of depolarising circuit noise and read-out errors which further supported the main finding of our work that the duplicate circuit offer superior performance towards mitigating of errors when compared to the QEC method. Our work reports a first assessment of the duplicate circuit approach for a quantum algorithm implementation and the documented evidence will pave the way for future scalable implementations of the duplicated circuit techniques for the error-mitigated practical applications of near-term quantum computers. 

\end{abstract}

\maketitle


\section{Introduction}

Computationally expensive problems such as breaking modern encryption algorithms or simulating quantum systems require exponentially large computing resources and quickly become intractable on classical computers when the problem size is increased. Contrarily, quantum computing, first envisioned by Fyenman \cite{feynman1982simulating}, is an emerging paradigm of computing, which is anticipated to be exponentially faster than a classical computer for certain hard problems which are classically intractable. For instance, Shor's factoring algorithm is a quantum solution designed to factor large integers which scales super-polynomially on classical computers \cite{365700}. Recently, two research groups have demonstrated the supremacy of quantum computers over classical supercomputers, first based on a superconducting quantum processor developed by Google \cite{arute2019quantum} and subsequently by using a photonic quantum computer developed by a Chinese research team \cite{zhong2020quantum}. The first experiment demonstrated by Google was to sample from the output of a random quantum circuit a million times, which would take a classical supercomputer 10,000 years, whereas, for a quantum computer with only 53 qubits, it required merely around 200 seconds to complete the task. The second experiment was to yield an output state space of the Gaussian boson sampling with around $10^{30}$ dimension within 200 seconds and its sample rate was around $10^{14}$ times faster than that of the state-of-the-art classical supercomputers. While these two demonstrations clearly establish the promise of immense computational capabilities of a quantum computer, a clear quantum advantage for practical applications is yet to be demonstrated.      

It is universally recognised that the primary limiting factor for today's quantum devices is noise or errors, which significantly reduce the accuracy of quantum algorithms. Even with error rates tracking below 1\%, their cumulative impact on large quantum circuits relevant for practical applications is detrimental to their capability to solve any real-world problems. Therefore, the development and testing of efficient error mitigation or correction strategies is crucial to fully unlock the promise of quantum computing. In this work, we focus on three quantum error mitigation strategies and compare their performance for experiments and quantum simulations implemented on IBM quantum system \cite{ibmquantum}. The first approach is based on encoding a noisy quantum state in a quantum error-detecting code to create error-mitigated logical qubits \cite{urbanek2020error}; the second approach uses entanglement between two (replicated) quantum states to calculate the error mitigated expectation value of the copied state \cite{huggins2021virtual}; the third approach iteratively optimizes the probability of measurement results of quantum circuits based on the Bayes' Theorem \cite{richardson1972bayesian, lucy1974iterative}. We tested these three methods on a noisy quantum computer and a quantum circuit simulator, respectively, using IBM quantum platform and performed a comprehensive comparative analysis based on the results. 

\begin{figure*}
\includegraphics[scale=0.4]{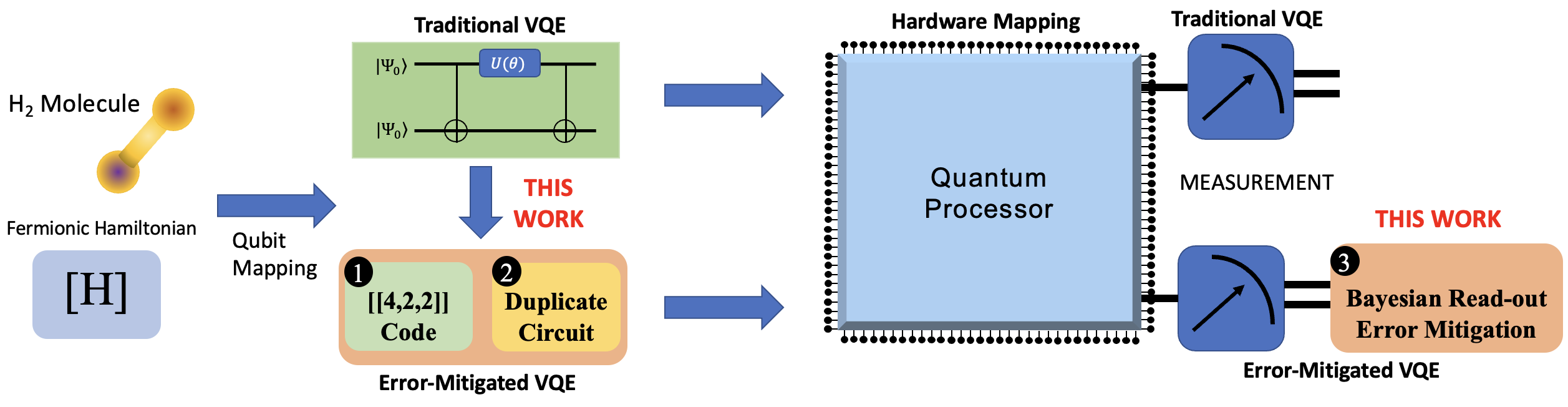}
\caption{A flowchart diagram which describes the procedure of our experiments. The fermionic Hamiltonian of H$_2$ molecule is mapped to two-qubit VQE circuit. In contrast to traditional approach where two-qubit circuit is mapped on a quantum processor, our work also implements three error mitigation schemes: \textcolor{white}{\protect \circled{1}} the [[4,2,2]] error-detecting code (QEC method), \textcolor{white}{\protect \circled{2}} the duplicate circuit and \textcolor{white}{\protect \circled{3}} the Bayesian read-out error mitigation (BREM). Whilst \textcolor{white}{\protect \circled{1}} and \textcolor{white}{\protect \circled{2}} mitigates the effect of circuit noise, \textcolor{white}{\protect \circled{3}} is implemented to tame the impact of measurement errors.}
\label{fig:all_procedures}
\end{figure*}

The benchmarking of the investigated error mitigation strategies is performed based on quantum chemistry simulations which is considered as one of the most promising applications of a quantum computer in near future \cite{bharti2022noisy}. Even for a simple molecule consisting of just a few atoms, the complexity of finding its ground state energy by solving multi-particle Hamiltonian would require tremendous computational power, surpassing the limits of classical computational techniques \cite{babbush2014adiabatic, aspuru2012photonic}. Variational quantum algorithms (VQA) such as variational quantum eigensolver (VQE) \cite{peruzzo2014variational} and quantum approximate optimization algorithm (QAOA) \cite{farhi2014quantum} have been demonstrated as the leading methods to find the ground state of molecules with limited quantum resources on today's quantum devices, solving problems in the field of quantum chemistry and quantum physics \cite{arute2020hartree, izmaylov2019unitary, kandala2017hardware, yung2014transistor, mcclean2016theory, li2017efficient, mcardle2019variational, barkoutsos2018quantum, dallaire2019low, romero2018strategies,grimsley2019adaptive}. In this work, we have implemented H$_2$ molecule VQE problem which allows us to perform a direct comparison of the three error mitigation schemes. Figure~\ref{fig:all_procedures} schematically illustrates the flowchart of our approach, highlighting three error mitigation techniques (labelled as \textcolor{white}{\protect \circled{1}} to \textcolor{white}{\protect \circled{3}}) tested in this work. While the traditional H$_2$ molecule VQE is solved on a quantum processor by optimisation of a two-qubit circuit, our work also implements 4-qubit [[4,2,2]] error-detecting code (QEC method) encoded VQE and the duplicate circuit VQE solutions. We also implemented Bayesian read-out error mitigation (BREM) to suppress the impact of measurement errors. Our results show that the duplicate circuit approach performs extremely well (and superior than the QEC method) against both the hardware noise as well as the simulated depolarisation noise. We also test multiple simultaneous implementations of the QEC and the duplicate circuit techniques which indicate significant impact of the cross-talk noise. Based on our results, we infer that a scalable implementation of duplicate circuit approach could lead to large-scale quantum chemistry simulations on NISQ devices, demonstrating the future direction of duplicate circuit. 

The remainder of the paper is divided in the following sections: Section II describes methods and the background literature. Section III reports our results obtained from both the experiments on the IBM quantum device, as well as from the quantum simulations with depolarisation noise. In Section IV, we discuss future directions and Section V provides the overall conclusions of our work.

\section{Methods and Background Literature}

\subsection{Variational Quantum Eigensolver}

The estimation of the ground state energy of a molecular system is an multi-particle electronic structure problem, which is one of the central problems in quantum chemistry. Due to limited capabilities of the current generation of quantum computers, the focus is on simulation of small-scale molecular systems which serve as a benchmark for quantum algorithms. A number of studies in the past have investigated the computation of the ground-state energy eigenvalue of H$_2$ molecule on a quantum computer or a quantum simulator using the well-known Variational Quantum Eigensolver (VQE) algorithm \cite{peruzzo2014variational, urbanek2020error, kandala2017hardware, colless2018computation, kandala2019error, o2016scalable}. The overview of VQA and VQE have been discussed in multiple review articles and interested readers are referred to the following references for details: \cite{cerezo2021variational, motta2021emerging, mcardle2020quantum, cao2019quantum}.

Here, we briefly summarise the VQE quantum circuit construction in the context of H$_2$ molecule. The VQE approach is a hybrid quantum-classical algorithm, which has been designed to calculate and optimize the ground state energy (the lowest eigenstate of a Hamiltonian) of a molecule efficiently \cite{kandala2017hardware, kokail2019self, mcclean2016theory, o2016scalable, peruzzo2014variational}. In this paper, we tested the performance of error mitigation techniques by implementing the VQE algorithm to find the ground-state energy of H$_2$ molecule.

We obtain the qubit Hamiltonian of H$_2$ molecule by following the approach in \cite{o2016scalable}. The H$_2$ Hamiltonian can be reduced from four qubits to two qubits. First, we apply the Born-Oppenhemier approximation \cite{born1927ann} to the second quantized form of the Hamiltonian of H$_2$ to assume the coordinates of two nuclei are fixed, choosing a minimal basis (we use STO-3G basis \cite{hehre1972self} in this paper), we can obtain a Hamiltonian in the form of:
\begin{equation}
\begin{aligned}\label{eq2quantizedhamiltonian}
H &= 
\sum_{i,j}f_{i,j}a^\dagger_i a_j
+\frac{1}{2}\sum_{i,j,r,s}f_{i,j,r,s}a^\dagger_i a^\dagger_j a_r a_s, 
\end{aligned}
\end{equation}
where
\begin{equation}
\begin{aligned}\label{eq2quantizedhamiltonian1}
f_{ij} &= 
\int dx\phi_i^*(x)
(-\frac{\nabla^2}{2}-\sum_I\frac{Z_I}{\abs{r-R_I}})\phi_j(x),
\\
f_{ijrs} &= 
\int dx_1dx_2
\frac{\phi_i^*(x_1)\phi_j^*(x_2)\phi_r(x_2)\phi_s(x_1)}{\abs{r_1-r_2}}.
\end{aligned}
\end{equation}
In this formalism, the one-body operator $\sum_{i,j}f_{i,j}a^\dagger_i a_j$ refers to moving a fermion from $jth$ spin orbital to $ith$ spin orbital, $f_{i,j}$ is the one-body integral, which illustrates the Coulomb interaction between electrons and nuclei, and the kinetic energies of the electrons; the two-body operator $\frac{1}{2}\sum_{i,j,r,s}f_{i,j,r,s}a^\dagger_i a^\dagger_j a_r a_s$ indicates that the operator moves two fermions from $rth$ and $sth$ spin orbitals to $ith$ and $jth$ spin orbitals, $f_{i,j,r,s}$ is the two-body integral, which represents the repulsion energy between the electrons \cite{mcardle2020quantum}. The Hamiltonian has an equal number of annihilation and creation operators, since the number of electrons in a molecular system is conserved.

Next, the Hamiltonian is mapped from fermions to qubits. This procedure can be achieved by using the Jordan-Wigner (JW) transformation \cite{jordan1993paulische} or the Bravyi-Kitaev (BK) transformation \cite{bravyi2002fermionic}. By applying the BK transformation, we can obtain a four-qubit Hamiltonian of the hydrogen molecule,
\begin{equation}
\begin{aligned}\label{eqh24qubithamiltonian}
H = &f_0 + 
f_1(Z_0 + Z_0Z_1) + 
f_2Z_1 \\ 
&+ f_3(Z_2 + Z_1Z_2Z_3) + 
f_4(Z_0Z_2 + Z_0Z_2Z_3) + 
f_5Z_1Z_3 \\ 
&+ f_6(X_0Z_1X_2 + Y_0Z_1Y_2 + X_0Z_1X_2Z_3 + Y_0Z_1Y_2Z_3) \\
&+ f_7(Z_0Z_1Z_2 + Z_0Z_1Z_2Z_3),
\end{aligned}
\end{equation}
where $\{f_i\}$ are real coefficients which depend on the bond length of the H$_2$ molecule and $\{X_i,Y_i,Z_i\}$ denote the corresponding Pauli operators acting on qubit $i$. Notice that qubit 1 and qubit 3 have only Pauli Z operators in the Hamiltonian, and our initial state has been set in a Hartree-Fock state \cite{mcardle2020quantum}. This means that these two qubits remain in the same state throughout the simulation, we can use this property to efficiently reduce the number of qubits required in our experiment. Finally, we further reduce our four-qubit Hamiltonian to two-qubit Hamiltonian by removing qubit 1 and qubit 3. The reduced qubit Hamiltonian of the hydrogen molecule acts on two qubits. These qubits and their corresponding coefficients can be re-indexed to form the following reduced Hamiltonian:
\begin{equation}
\begin{aligned}\label{eqh2hamiltonian}
H = g_0 + g_1Z_1 + g_2Z_2 + g_3Z_1Z_2 + g_4Y_1Y_2 + g_5X_1X_2,
\end{aligned}
\end{equation}
where $\{g_i\}$ are the coefficients calculated classically, which vary based on different bond lengths of the H$_2$ molecule. In this paper, we use values of $g_i$ published in Ref.~[\onlinecite{colless2018computation}]. The expectation value $\left\langle{H}\right\rangle$ of the Hamiltonian is calculated as
\begin{equation}
\begin{aligned}\label{eqexpvalueh2}
\left\langle{H}\right\rangle &= \bra{\psi(\theta)}H\ket{\psi(\theta)} \\
&= \sum_{i}g_i\bra{\psi(\theta)}\prod_j\sigma^j_i\ket{\psi(\theta)} \\
&= g_1 + 
g_2\left\langle{Z_1}\right\rangle + 
g_3\left\langle{Z_2}\right\rangle\\
&+ g_4\left\langle{Z_1Z_2}\right\rangle + 
g_5\left\langle{Y_1Y_2}\right\rangle +
g_6\left\langle{X_1X_2}\right\rangle.
\end{aligned}
\end{equation}
The variation principle of quantum mechanics implies that, $\left\langle{H}\right\rangle \ge \varepsilon_0$ where $\varepsilon_0$ is the exact ground state energy of the H$_2$ molecule.

We prepared the corresponding parametrized quantum circuits based on the simplified two-qubit Hamiltonian and to form the Ansatz state,
\begin{equation}
\begin{aligned}\label{ansatzstate}
\ket{\Psi(\theta)},
\end{aligned}
\end{equation}
where $\theta$ is the parameter that will be used in VQE circuit. Then we repeatedly measured each term in the Hamiltonian on a quantum processor to estimate the ground state energy of H$_2$. The trial wavefunction $\ket{\Psi(\theta)}$ in $\bra{\Psi(\theta)}H\ket{\Psi(\theta)}$ that we used is the UCC ansatz \cite{bartlett1989alternative}, which has been calculated in Ref.~[\onlinecite{urbanek2020error}] and given by
\begin{equation}
\begin{aligned}\label{equcch2}
\ket{\Psi(\theta)} = e^{i(-\theta)Y_1X_2}\ket{\Phi},
\end{aligned}
\end{equation}
where in this paper, we use $\ket{\Phi} = \ket{00}$ as the reference state, which is obtained from the Hartree-Fock method.

\begin{figure*}
\includegraphics[scale=0.55]{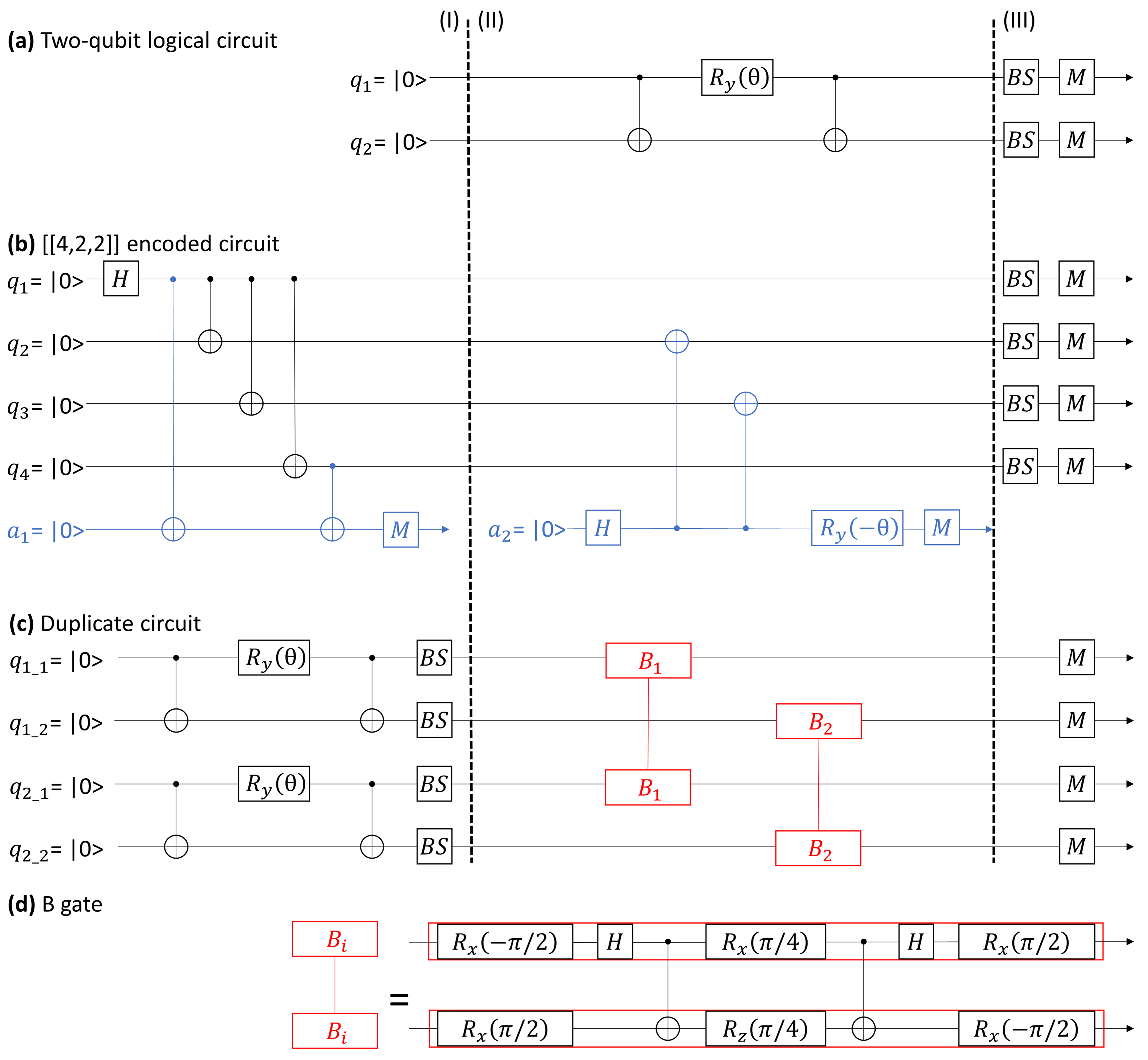}
\caption[Quantum circuits for the three VQE implementations.]{Quantum circuits are plotted for the [[4,2,2]] quantum error-detecting code (QEC), duplicate circuit and Bayesian read-out error mitigation scheme. BS gates denote basis transformations based on the corresponding measured term in qubit Hamiltonian and M gates measure the outputs of qubits. (a) The diagram of a two-qubit circuit based on trial wavefunction defined by Eq.~\ref{eq422code}. The circuit in section I prepares the Hartree-Fock initial state $\ket{00}$ as the initial state. The circuit in section II performs the UCC exponential in the trial wavefunction. Measurements are applied in section III. (b) The diagram of the QEC method based on the [[4,2,2]] quantum error-detecting code in Eq.~\ref{eq422code} and trial wavefunction in Eq.~\ref{equcch2}. In section I we prepare the logical initial state $\ket{00}_L$, ancilla qubit $a_1$ is used to detect if errors occur during the state preparation. In section II, ancilla qubit $a_2$ is used to perform the UCC exponential. In section III we apply the corresponding basis transformations and measure the outputs. (c) The diagram of the duplicate circuit. In section I, we implement two two-qubit circuits on the first two qubits and the last two qubits respectively, $q_{i_j}$ refers to the $jth$ qubit in $ith$ subsystem. Basis transformations have been applied in this section as well. The circuit in section II is a single layer of B gates with matrix form defined by Eq.~\ref{eqbgate}, $B_i$ denotes B gate acts on $ith$ qubit in each subsystem. The circuit in section III measures the outputs. (d) A realization of B gate on $ith$ qubit of two subsystems.}
\label{fig:all_circs}
\end{figure*}

\subsection{Depolarizing Error Model}

The depolarizing error, which is also known as a depolarizing channel, implements noise on single-qubit and is defined as
\begin{equation}
\begin{aligned}\label{eqdepolarizing1}
E(\rho) &= p\rho_{cms}+(1-p)\rho, \\
\rho_{cms} &= \frac{I}{2^n} = 
\frac{1}{2^n}\sum_{i=1}^{2^n}\ket{\psi_i}\bra{\psi_i},
\end{aligned}
\end{equation}
where for a single qubit, with probability $p$, that qubit is depolarized to a completely mixed state $\rho_{cms} = I/2$, and with probability $1-p$ the qubit stays untouched \cite{nielsen2002quantum}. More generally, in Eq.~\ref{eqdepolarizing1}, $n$ is the number of qubits the depolarizing channel acts on.

\subsection{Quantum Error-Detecting Code}

With the increasing size of quantum computers, mitigating and correcting the effect of noise is becoming more and more vital, which means quantum circuits should be more noise resilient, to allow Noisy Intermediate-Scale Quantum (NISQ) devices \cite{preskill2018quantum} to achieve greater performance. The noise can come from a variety of sources, most fundamental of which is decoherence induced by the interaction between qubits and the environment and leads to errors occurring on the quantum computer. To mitigate the effect of noise or errors, researchers have developed quantum error correction and detection codes to try to correct induced errors \cite{babbush2014adiabatic, fowler2009high, vaidman1996error, gottesman2010introduction}. Here we explore the performance of a QEC method \cite{grassl1997codes, gottesman2016quantum, urbanek2020error, linke2017fault} to encode four physical qubits to two logical qubits. The QEC has the stabilizers $Z_4Z_3Z_2Z_1$ and $X_4X_3X_2X_1$, which can detect any single-qubit error in the encoded state. The distance between any two logical states is two. The logical code words are:
\begin{equation}
\begin{aligned}\label{eq422code}
\ket{00}_L &\to \frac{1}{\sqrt{2}}(\ket{0000}+\ket{1111}), \\
\ket{01}_L &\to \frac{1}{\sqrt{2}}(\ket{0011}+\ket{1100}), \\
\ket{10}_L &\to \frac{1}{\sqrt{2}}(\ket{0101}+\ket{1010}), \\
\ket{11}_L &\to \frac{1}{\sqrt{2}}(\ket{0110}+\ket{1001}), 
\end{aligned}
\end{equation}
where the first qubit refers to the rightmost qubit in both logical and physical states. 

There is no need to use ancilla qubits to perform logical operations on a single logical states. Some operators on single logical qubit are the product of single physical qubit Pauli operators. For instance, the logical $X$ operator acting on the first logical qubit is $I_4I_3X_2X_1$ and the logical $X$ operator acting on the second logical qubit is $I_4X_3I_2X_1$. $I_4Z_3I_2Z_1$ is the logical $Z$ operator applied to the first logical qubit and $I_4I_3Z_2Z_1$ is the logical $Z$ operator applied to the second logical qubit. The logical $H$ operation acting on both logical qubits is $H_4H_3H_2H_1$. For logical $CNOT$ gate, which performs $CNOT_{L,12}=SWAP_{13}$, where $CNOT_{L,12}$ denotes control qubit on the first logical qubit and target qubit on the second logical qubit, similarly, $CNOT_{L,21}=SWAP_{12}$. The implementations of the two-qubit circuit to prepare the trial state, and the equivalent logical circuit for the QEC are shown in (a) and (b) in Fig.~\ref{fig:all_circs} respectively.

\subsection{Duplicate Circuit}

The duplicate circuit method is a second technique to mitigate the effect of errors in a quantum circuit. First proposed by Cotler \textit{et al} \cite{cotler2019quantum}, it makes use of multiple copies of a circuit being performed on a quantum computing device and makes correlated measurements between these in an attempt to reduce the resulting error. We also note that another study by Huggins \textit{et al} \cite{huggins2021virtual} translated a similar approach from bosonic systems to qubit systems. We apply this approach to our work by connecting two identical two-qubit circuits (hence named "duplicate circuit" in this paper) with a layer of two-qubit gates and measure the qubits in computational basis.

We calculate the expectation value of duplicate circuit according to 
\begin{equation}
\begin{aligned}\label{eqexpdup}
E(O) =
\frac{\Tr(O\rho^N)}{\Tr(\rho^N)} = 
\frac{\Tr(O^{(N)}S^{(N)}\rho^{\otimes N})}{\Tr(S^{(N)}\rho^{\otimes N})},
\end{aligned}
\end{equation}
where $\rho$ denotes a density matrix of the quantum system, in our case, $\rho$ is a $4\times4$ matrix generated from the two-qubit circuit, which can be diagonalised to $\rho = \sum p_i \ket{\phi_i}\bra{\phi_i}$. Raising $\rho$ to the $N$ power gives,
\begin{equation}
\begin{aligned}\label{eqrhois}
\rho^N = \sum_ip^N_i\ket{\phi_i}\bra{\phi_i},
\end{aligned}
\end{equation}
where $\ket{\phi_i}$ are the four eigenstates of the original density matrix, $\rho$. Under exponentiation, the dominant eigenvalue is amplified exponentially in $N$, whereas, in contrast, the non-dominant eigenvalues are suppressed exponentially in $N$. $O^{(N)}$ represents the observable operator $O$ that acts on $N$ quantum systems, 
\begin{equation}
\begin{aligned}\label{eqOoperator}
O^{(N)} &= \frac{1}{N}\sum_{i=1}^NO^i, \\
O^i &= I^{\otimes M}_1 \otimes I^{\otimes M}_2 \otimes \dots \otimes O^{\otimes M}_i \otimes \dots \otimes I^{\otimes M}_N, 
\end{aligned}
\end{equation}
where each system has $M$ qubits. In our case, $O$ is the Pauli $Z$ operator acting on two copied systems, namely
\begin{equation}
\begin{aligned}\label{eqOisZ}
Z^{(2)} = \frac{1}{2}(Z^1+Z^2).
\end{aligned}
\end{equation}
$S^{(N)}$ defined as the cyclic shift operator acts on $N$ quantum systems,
\begin{equation}
\begin{aligned}\label{eqSisCyclic}
S^{(N)}\ket{\psi_1\psi_2\dots\psi_N} = \ket{\psi_2\psi_3\dots\psi_1}.
\end{aligned}
\end{equation}
In our case, we have two copied systems in the duplicate circuit, hence $S^{(2)}$ works the same as a quantum SWAP gate. In order to simultaneously diagonalize $S^{(2)}$ and $Z^{(2)}$ which act on the pairs of qubits independently as the factorized tensor product of operators, Huggins et al \cite{huggins2021virtual} defined a two-qubit unitary gate called beam-splitter unitary gate ($B$ gate),
\begin{equation}
\begin{aligned}\label{eqbgate}
B^{(2)} &= B^{(2)}_1 \otimes B^{(2)}_2 \otimes \dots \otimes B^{(2)}_M, \\
B_i &=
    \begin{bmatrix}
    1 & 0 & 0 & 0\\
    0 & \frac{\sqrt{2}}{2} & -\frac{\sqrt{2}}{2} & 0\\
    0 & \frac{\sqrt{2}}{2} & \frac{\sqrt{2}}{2} & 0 \\
    0 & 0 & 0 & 1
    \end{bmatrix},
\end{aligned}
\end{equation}
where $M$ is the total number of qubits in each system and the index $i$, refers to the $ith$ duplicate system. An implementation of the duplicate circuit using two two-qubit logical circuits is shown in Fig.~\ref{fig:all_circs}(c) and a realization of $B$ gate is shown in Fig.~\ref{fig:all_circs}(d).

\begin{SCfigure*}
\includegraphics[scale=0.35]{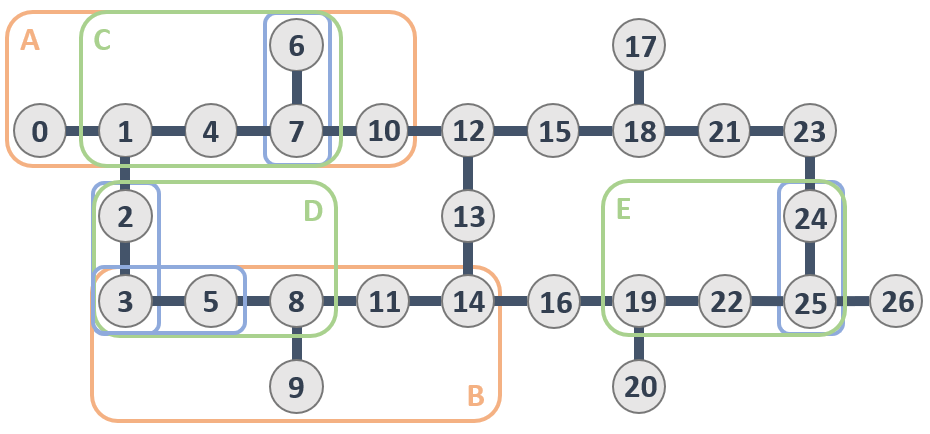}
\caption{Chip geometry of $ibmq\_toronto$. Numbers represent the index of qubits. Blue boxes circle the qubits that are mapped by two-qubit circuit, orange boxes circle the qubits that are mapped by the QEC method and green boxes circled the qubits that are mapped by duplicate circuit.}
\label{fig:toronto}
\end{SCfigure*} 

\subsection{Bayesian Read-Out Error Mitigation}

In this work, we implemented Bayesian read-out error mitigation (BREM) (also known as Bayesian unfolding) to mitigate the impact of read-out error \cite{urbanek2020error, nachman2020unfolding}. read-out error is the error that occurs in measurements, where the state of a qubit is incorrectly determined. That is, a qubit in state $\ket{0}$ is measured as $\ket{1}$ and vice versa. The BREM approach \cite{richardson1972bayesian, lucy1974iterative} is an iterative technique based on Bayes' Theorem to correct read-out errors. The recursive equation used is:
\begin{equation}
\begin{aligned}\label{eqbayesian}
t_i^{r+1} = 
\sum_j\frac{Pr(\mathrm{measure}\ j|\mathrm{truth\ is}\ i)t_i^r}
{\sum_sPr(\mathrm{measure}\ j|\mathrm{truth\ is}\ s)t_s^r}
Pr(\mathrm{truth\ is}\ j),
\end{aligned}
\end{equation}
where $r$ is the iteration number, $t_i^0$ is called a prior truth spectrum, we set it as a uniform distribution. $Pr(\mathrm{measure}\ j|\mathrm{truth\ is}\ i)$ is a response matrix, which is the probability of measuring state $j$ when given the state $i$. The response matrix can be determined experimentally by constructing $2^n$ calibration circuits, where $n$ is the number of qubits, each circuit is constructed with only  Pauli $X$ gates and measured on $Z$ basis. $Pr(truth\ is\ j)$ is the measured spectrum, which is the probability of measuring state $j$ when given the state $j$. $i$, $j$ and $s$ are the number of states, which is $2^n$.

\section{Implementation on IBM System}

We summarise our VQE experiments as follows. We calculate the expectation value of four terms: $\left\langle{Z_1}\right\rangle$, $\left\langle{Z_2}\right\rangle$, $\left\langle{Z_1Z_2}\right\rangle$, and $\left\langle{X_1X_2}\right\rangle$ independently in our qubit Hamiltonian on an IBM quantum device by using three circuits shown in Figure~\ref{fig:all_circs}. According to the Pauli operators in these four terms, we apply the corresponding basis transformation to measure them on the correct basis. To be specific, for $\left\langle{Z_1}\right\rangle$, $\left\langle{Z_2}\right\rangle$, and $\left\langle{Z_1Z_2}\right\rangle$ terms, no basis transformation needs to be applied; for $\left\langle{X_1X_2}\right\rangle$ term, Hadamard gates will be used as the basis transformation gates to transform the basis from Z basis to X basis. For the values of the variational parameter $\theta$, we choose 257 values of $\theta \in [-\pi, \pi]$. Since we have only one parameter, all of the 257 values will be implemented for all three circuits. We sample each expectation value with 8192 shots for each value of the variational parameter. For inter-atomic distance between electrons in H$_2$, we choose 78 values of $r \in [0.1, 3.95]\ \mathring{\mathrm{A}}$, ground state energy for each separation is estimated by determining the minimum energy $\left\langle{E}\right\rangle_\theta$ among each of the 257 angles. We will compare each of the methods described above to determine their effectiveness at mitigating errors present in the devices for this VQE experiment.

\subsection{Quantum Hardware Experiments}

We executed three circuits on $ibmq\_toronto$, which is one of the 27-qubit IBM Quantum Falcon Processors \cite{ibmquantum} (Geometry shown in Fig.~\ref{fig:toronto}), to test the performance for each mitigation technique applied to the VQE algorithm. Since the noise level of each qubit varies over time, we arbitrarily choose the qubits to implement these circuits. For each circuit, we first used the measured outputs of the circuit to calculate the raw ground state energy of H$_2$ molecule. Then we calculated the optimized ground state energy by using the BREM technique described earlier in section II E, which can be obtained by applying the BREM approach to the raw measurement results to mitigate the effect of read-out error. For the QEC method, we then discard the states for which an error was detected and renormalise the remaining states.

\subsubsection{Two-Qubit Circuit}

\begin{figure*}
\includegraphics[scale=0.6]{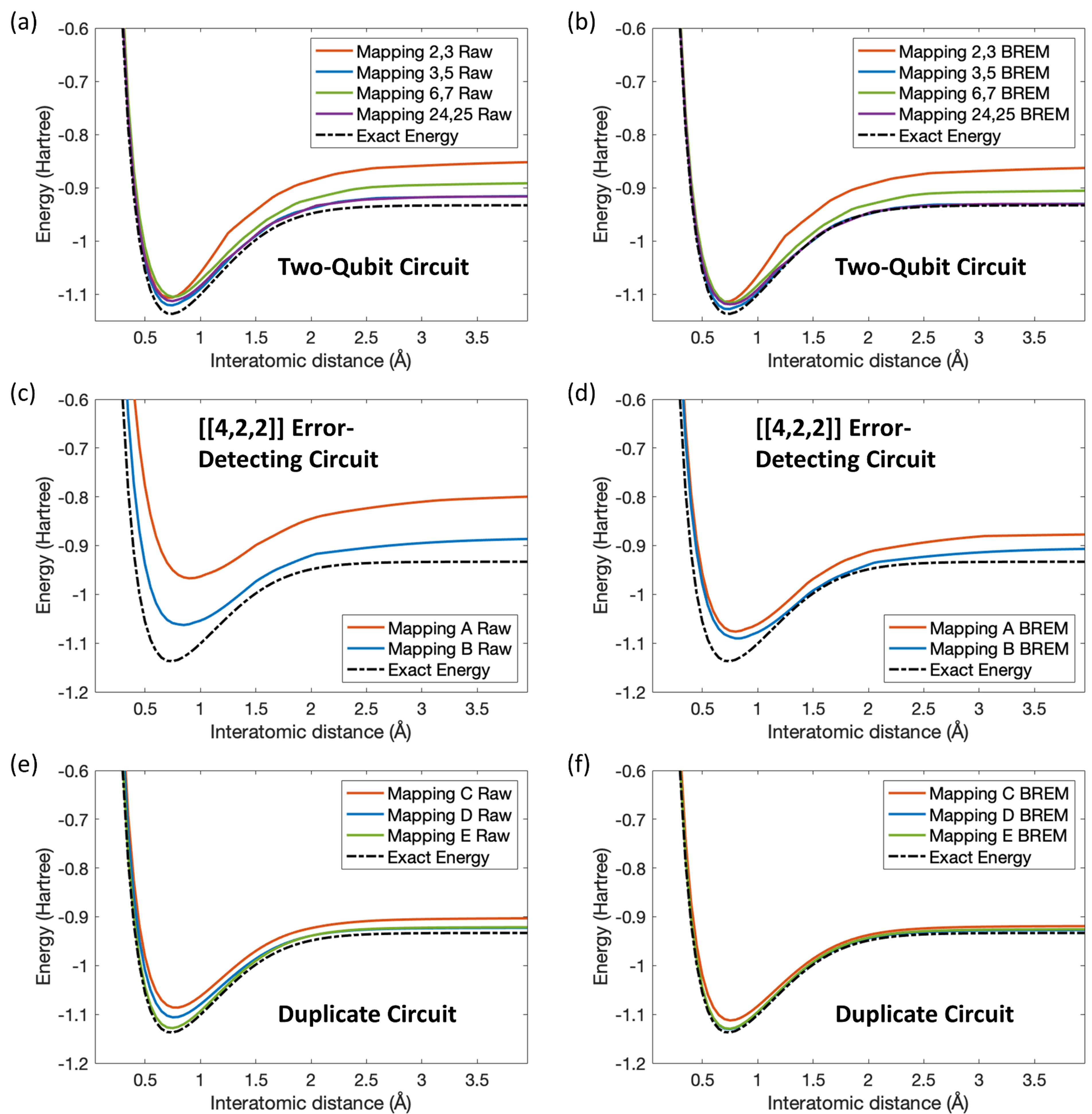}
\caption{The comparison of experimental results of two-qubit circuit, the [[4,2,2]] quantum error-detecting code (QEC) circuit and duplicate circuit, where Raw in legends for all figures denotes the values calculated from raw data and BREM in legends denotes the application of the Bayesian read-out error mitigation technique, which shows the optimized energies. (a) and (b) illustrate the results from two-qubit experiments, Mapping i,j in legends denotes the indices of qubits that are circled by blue boxes in Fig.~\ref{fig:toronto}; (c) and (d) show the results of the QEC method, Mappings A and B in legends denote qubits that are circled by orange boxes with the corresponding colored labels in Fig.~\ref{fig:toronto}, respectively; (e) and (f) represent the results of duplicate circuit, Mappings C, D and E in legends denote qubits that are circled by green boxes with the corresponding colored labels in Fig.~\ref{fig:toronto}, respectively.
(a) Raw ground state energy of four mappings for two-qubit circuit. 
(b) Optimized ground state energy of four mappings for two-qubit circuit.
(c) Raw ground state energy of two mappings for the QEC method.
(d) Optimized ground state energy of two mappings for the QEC method.
(e) Raw ground state energy of three mappings for the duplicate circuit.
(f) Optimized ground state energy of three mappings for the duplicate circuit. }
\label{fig:all_toronto1}
\end{figure*}

\begin{figure*}
\includegraphics[scale=0.6]{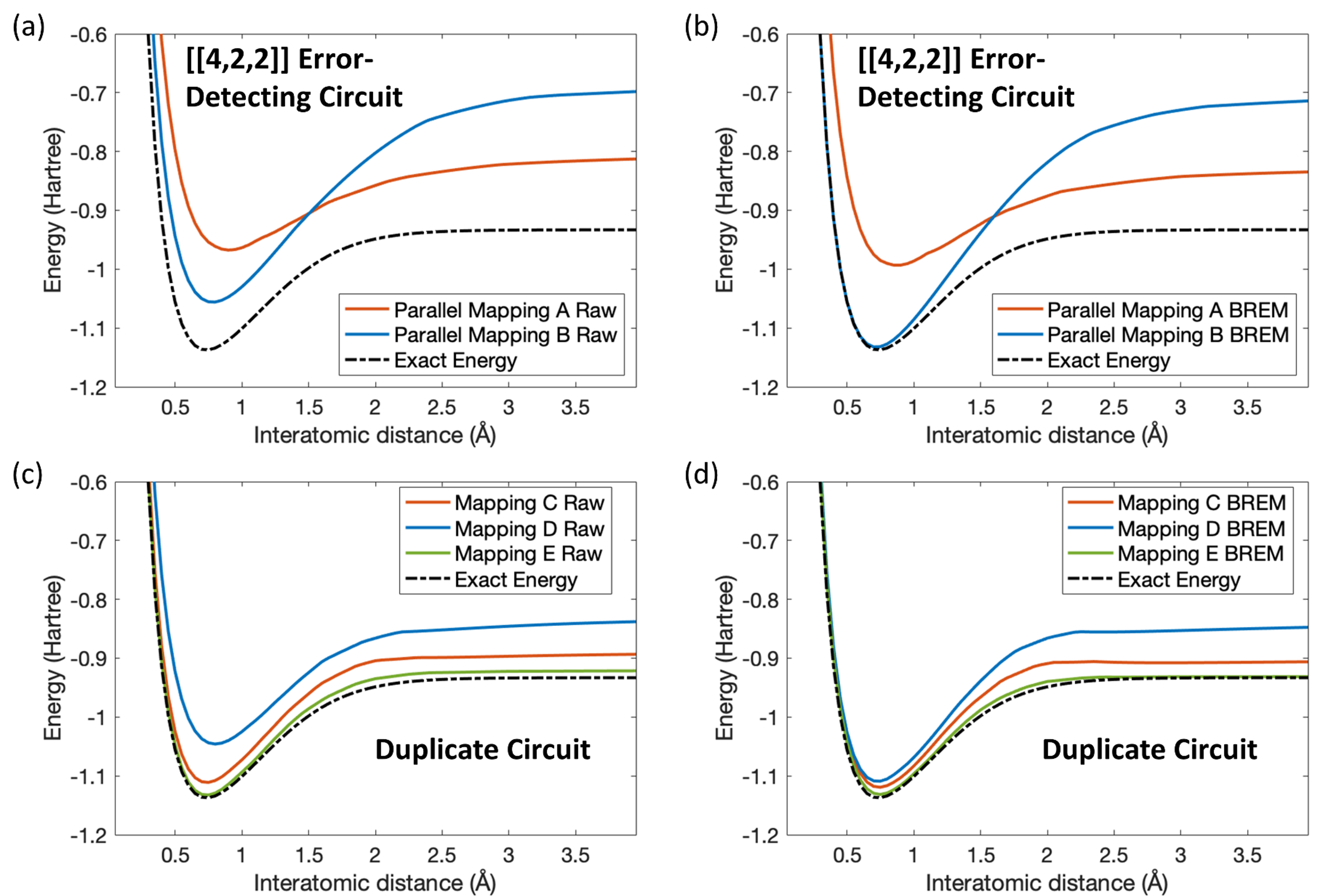}
\caption{The comparison of experimental results of the [[4,2,2]] quantum error-detecting code (QEC) circuit and duplicate circuit for parallel (simultaneous) execution of two mappings and three mappings, respectively. Mapping A and Mapping B in (a) and (b) denote qubits that are circled by orange boxes in Fig.~\ref{fig:toronto} with the corresponding colored labels, respectively; Mapping C, Mapping D and Mapping E in (c) and (d) denote qubits that are circled by green boxes in Fig.~\ref{fig:toronto} with the corresponding coloured labels, respectively. 
(a) The raw experimental results of two mappings of the QEC method with parallel execution. 
(b) The experimental results of two mappings of the QEC method with parallel execution after the application of BREM scheme.
(c) The raw experimental results of three mappings of duplicate circuit with parallel execution. 
(d) The experimental results of three mappings of duplicate circuit with parallel execution after the application of BREM scheme.}
\label{fig:all_toronto2}
\end{figure*}

We implemented the two-qubit circuit with four mappings onto $ibmq\_toronto$ (see Figure~\ref{fig:toronto}) and the VQE experimental results are shown in Fig.~\ref{fig:all_toronto1}(a) and Fig.~\ref{fig:all_toronto1}(b). The BREM improves the fidelity of the measured VQE results in each mapping as expected. The results for each of the four mappings investigated vary due to the different noise level on the mapped qubits, both raw ground state energy curve and optimized ground state energy curve with mapping. The mapping utilizing qubits three and five had have the best performance among all raw energy curves and all optimized energy curves, which indicate qubit 3 and qubit 5 have the lowest noise level among all mapped qubits during the execution. Contrarily, the mapping utilizing qubits two and three had exhibits least agreement with the exact energy calculation and in the next two sections, we will explore if better results can be achieved by using error-mitigation strategies.

\subsubsection{[[4,2,2]] Error-Detecting Circuit}

We tested two mappings (A and B, see Figure~\ref{fig:toronto}) of the QEC method under two different scenarios on $ibmq\_toronto$. In the first scenario, we tested the performance of two mappings by executing them independently (or serially), \textit{i.e.}, we first implemented VQE with mapping A, and then we implemented VQE with mapping B (refer to Fig.~\ref{fig:toronto}). For the second scenario, we tested the performance of the two mappings by executing them in parallel, which is, we apply mapping A and mapping B simultaneously, measuring the outputs of the two mappings at the same time at the stage of obtaining the raw data in VQE, remaining other procedures of VQE for each mapping the same. 

The VQE results of the first scenario are shown in Fig.~\ref{fig:all_toronto1}(c) and Fig.~\ref{fig:all_toronto1}(d). These results show that mapping B performs better than mapping A in both raw ground state energy and optimized ground state energy. The curve for raw ground state energy of mapping B is closer to the exact energy curve than the curve for optimized ground state energy when the inter-atomic distance is larger than around 1.25 $\mathring{\mathrm{A}}$, which indicates the choice of mapping qubits is vital which can further improves the fidelity of VQE results.

The experimental results of the second scenario are represented in Fig.~\ref{fig:all_toronto2}(a) and Fig.~\ref{fig:all_toronto2}(b). In this case, both mapping A and mapping B generate highly noisy results, since there are twelve qubits used during the experiment, whereas an individual mapping requires only six qubits. More qubits induce a larger error on a noisy quantum computer. One possible explanation for this is that cross-talk error is present. Cross-talk error is generally referred to the unwanted coupling generated between signal paths \cite{mazda2014telecommunications}, which commonly occurs between qubits on noisy quantum devices. In scenario two, the effect of the cross-talk errors is stronger than in scenario one since the qubits in two mappings are close to each other according to the geometry of the quantum chip as shown in Fig.~\ref{fig:toronto}. This causes enhanced inaccuracy in the experimental results for both mappings.

Comparing the experimental results from the two scenarios for the QEC method to the experimental results of the two-qubit circuit, our results show that both of the scenarios perform worse than the two-qubit circuit in the VQE experiments. The QEC method consists of six qubits, each of the qubits increases the probability of errors occurring during the execution, so it is naturally noisier than the two-qubit circuit. In addition, the two-qubit circuit can be directly mapped on the corresponding qubits, whereas larger circuits like the QEC method requires auxiliary gates, such as SWAP gate, to map the circuit on the corresponding qubits. This procedure also adds noise to the circuit. Overall, the mapping structures of quantum circuits and the size of quantum circuits both affect the experimental results. We note that our results are in contrast to an earlier study using IBM Sydney machine \cite{urbanek2020error}, where QEC method was shown to perform better than two-qubit mapping. We believe that this could be due to different noise in IBM Sydney and IBM Toronto machines. We could not repeat our experiments on IBM Sydney machine which is no longer available from IBM Quantum platform.

\subsubsection{Duplicate Circuit}

In this section, we analyse a second error-mitigation scheme (duplicate circuit) which is described earlier in Section II D. For this case, we considered three 4-qubit mappings (C, D, and E) of the duplicate circuit onto the $ibmq\_toronto$ processor as illustrated in Fig.~\ref{fig:toronto}. We again consider two scenarios as discussed in the last section. For the first scenario, we tested the performance of three mappings by executing them independently (one by one). For the second scenario, we tested the performance of the three mappings by executing all of them in parallel, which is, we apply mapping C, D and E simultaneously, measuring the outputs of three mappings at the same time at the stage of obtaining the raw data in VQE, remaining other procedures of the VQE implementation for each mapping the same. 

The comparison of experimental results for duplicate circuit implementations under scenario one is shown in Fig.~\ref{fig:all_toronto1}(e) and Fig.~\ref{fig:all_toronto1}(f). We find that all three mappings exhibit high fidelity results, performing better than the two-qubit and QEC method results. The duplicate circuit consists of many more gates where each gate adds additional noise. In theory, it is easier to obtain an accurate expectation value of the duplicated system when the number of copied systems $N$ in Eq.~\ref{eqexpdup} is larger, but in practice, more gates introduce more noise to the quantum system. Thus, there is a trade-off of the advantage conferred by having more duplicate systems, and the additional noise that increasingly larger depth $B$ gates add. As the results for duplicate circuit indicate very close agreement with the exact energy, it is concluded that the duplicate circuit provides significant advantage over the QEC method in mitigating hardware errors. 

The VQE results of scenario two are shown in Fig.~\ref{fig:all_toronto2}(c) and Fig.~\ref{fig:all_toronto2}(d), where all three duplicate circuit mappings are executed concurrently. Mapping E generated the highest fidelity results in both raw experimental data and after the application of BREM approach. Overall our results show that while cross-talk impacts duplicate circuit performance, generally the duplicate circuit scheme was still observed to be quite effective in mitigating this kind of complex noise mechanism.

\begin{figure*}[h!]
\centering
\includegraphics[scale = 0.6]{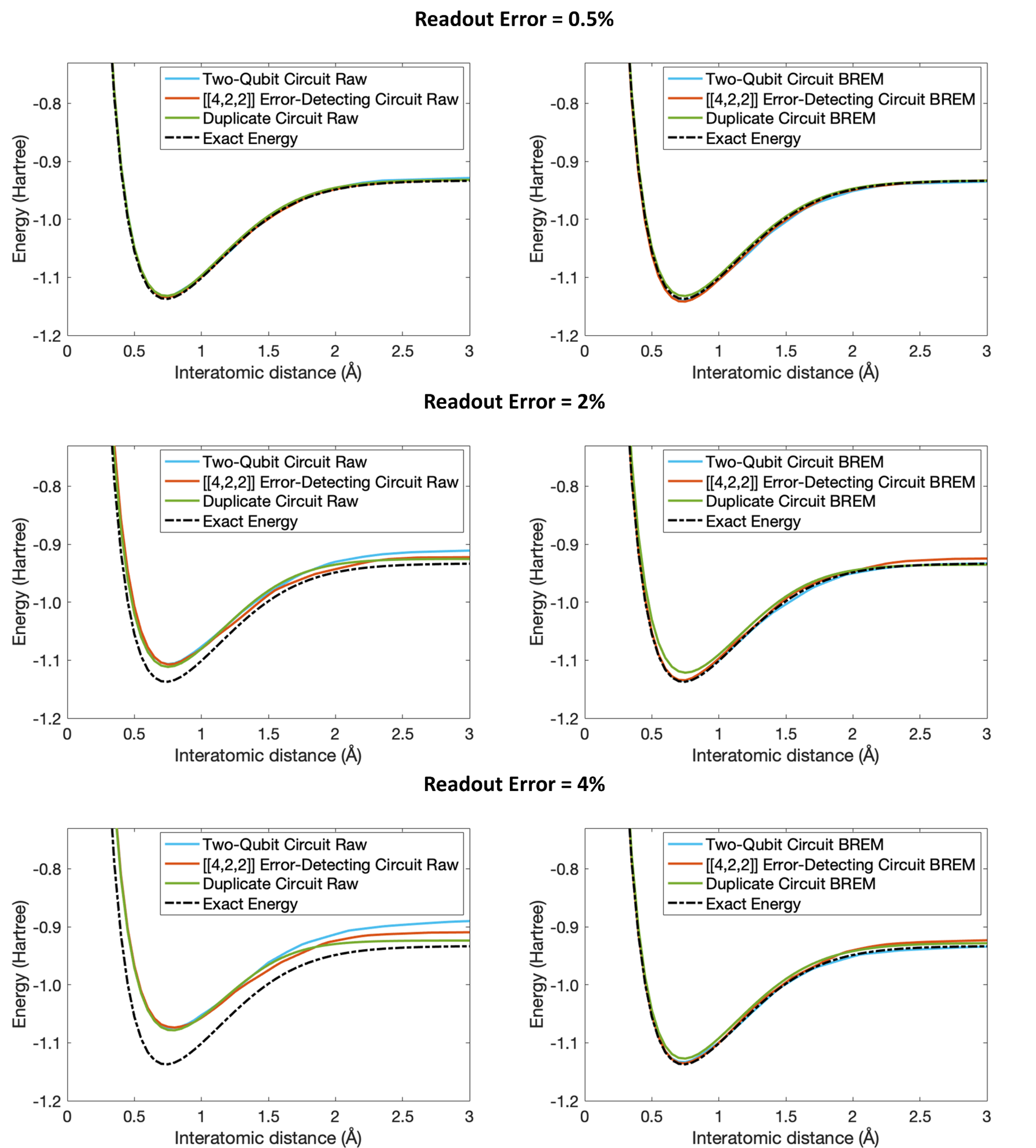}
\caption[VQE Results on read-out error Simulator]{The comparison of VQE results of three circuits on the simulator with read-out error rates 0.5\%, 2\% and 4\%.}
\label{fig:allreadout}
\end{figure*}

\subsection{Analysis of Noise Interplay based on Quantum Simulations}

In the previous section, the results were reported from experiments on IBM quantum devices which include hardware noise. However, in those experiments, it was not possible to individually understand the impact of read-out error and circuit noise. In this section, our aim is to test and compare the performance of the BREM technique for the two-qubit circuit, the QEC method and the duplicate circuit on the AerSimulator (a noisy quantum circuit simulator in qiskit \cite{qiskit}) under three different noisy simulation scenarios. The simulator in each scenario included varying types and degrees of noise. To be specific, for the first scenario, we added only read-out error to the simulator; for the second scenario, we added only depolarizing error to the simulator; and for the third scenario, we added both read-out error and depolarizing error to the simulator. This analysis helps us to understand the relative impact of each type of noise on the VQE implementation and the comparative performance of the investigated error mitigation techniques.

\subsubsection{read-out errors}

In first set of simulations, we added only read-out errors to the noisy simulator, that is, only read-out errors can occur upon the measurements during the simulations (all quantum gates were assumed to be noise-free). We tested the performance of the three circuits (Figure~\ref{fig:all_circs}) under three read-out error levels, 0.5\%, 2\% and 4\% respectively. The corresponding results are plotted in Figure~\ref{fig:allreadout}. The read-out error can be regarded as a conditional probabilistic error since it can be summarized as $P(\mathrm{measure} \ket{0}|\mathrm{truth\ is} \ket{0})$, $P(\mathrm{measure} \ket{1}|\mathrm{truth\ is} \ket{0})$ given input state $\ket{0}$ and $P(\mathrm{measure} \ket{0}|\mathrm{truth\ is} \ket{1})$, $P(\mathrm{measure} \ket{1}|\mathrm{truth\ is} \ket{1})$ given input state $\ket{1}$ to the measured qubit, matching the conditions of the BREM technique to iteratively solve the problems based on Bayes' Theorem. Therefore, read-out errors can be largely mitigated by BREM approach on quantum circuits. 

\begin{figure*}[h!]
\centering
\includegraphics[scale = 0.6]{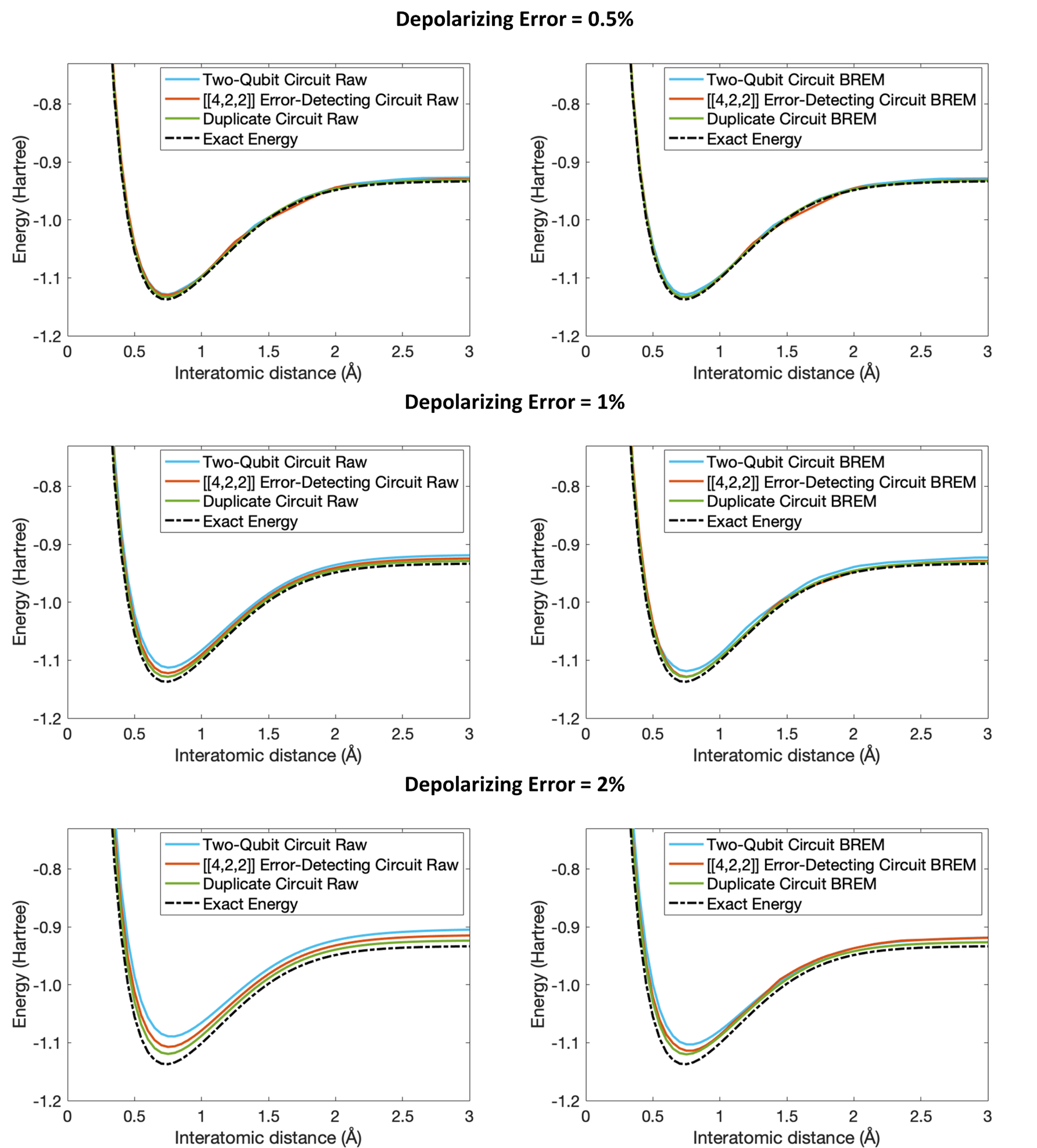}
\caption[VQE Results on Depolarizing Error Simulator]{The comparison of VQE results of three circuits on the simulator with depolarizing error rates 0.1\%, 1\% and 2\%.}
\label{fig:alldep}
\end{figure*}

\begin{figure*}[tbh!]
\centering
\includegraphics[scale = 0.35]{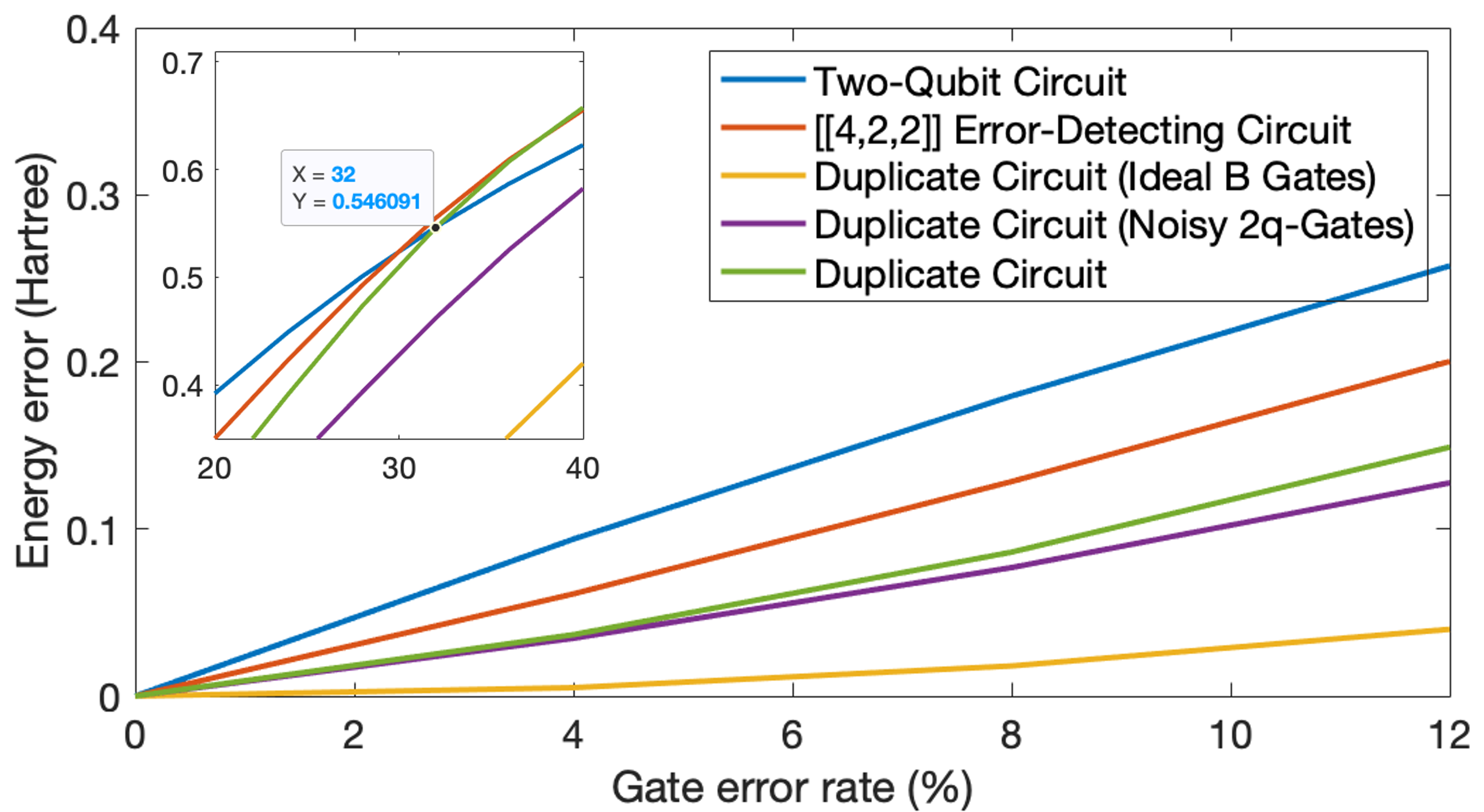}
\caption[Energy error of five circuits with depolarizing error model]{Energy error of five circuits with depolarizing error model in $\SI{0.75}{\angstrom}$ inter-atomic distance. The [[4,2,2]] error-detecting circuit (QEC method) and duplicate circuit perform better than two-qubit circuit when gate error rate is lower than around $30\%$, the duplicate circuit with ideal B gate shows the best performance out of the five circuits. Note that these simulations do not include read-out error or BREM implementation.}
\label{fig:energy_errors}
\end{figure*}

The computed ground state energy plots are shown in Fig.~\ref{fig:allreadout}. These results show that for each read-out error level, the BREM approach can indeed significantly improve the fidelity of the estimated ground state energy of the H$_2$ molecule in two-qubit circuits and the QEC method -- the corresponding optimized curves are very close to the exact energy. Our results also indicate that the performance of the BREM technique is approximately similar for all three circuit implementations, roughly independent of read-out noise level. Therefore, we conclude that BREM could in general improve quantum circuit fidelities, irrespective of the other error mitigation strategies implemented to cancel the impact of circuit noise.  

\subsubsection{Depolarizing Noise}

In these second set of simulations, we implemented three circuits (Figure~\ref{fig:all_circs}) on a noisy simulator deployed with only depolarizing error model (see Section II-B  for details), whereas the read-out error was assumed to be zero. Since we only use single-qubit gates and two-qubit gates in three circuits, based on the definition above, we added single-qubit depolarizing noise and two-qubit depolarizing noise to the simulator at the same rate.

The simulation results are shown in Fig.~\ref{fig:alldep}, for a variety of different depolarizing noise rates (0.5\%, 1\%, and 2\%). In the absence of any explicit read-out error, it is apparent that the BREM technique barely mitigates the effect of depolarizing error for three circuits; all three optimized curves are very close to the corresponding raw curves, since the depolarizing error can occur multiple times on each of qubits. For instance, when implementing a one-qubit circuit with multiple single-qubit gates, each gate has a probability $p$ to occur a depolarizing error, according to the BREM approach, $Pr(\mathrm{truth\ is}\ j)$ in Eq.~\ref{eqbayesian} is hard to be the corresponding probability to the likelihood $Pr(\mathrm{measure}\ j|\mathrm{truth\ is}\ i)$, which has the different depolarizing error distribution on the gates. Whereas for the circuit with only read-out errors, which occur only at the stage of measurement, so $Pr(\mathrm{truth\ is}\ j)$ has the corresponding probability distribution to $Pr(\mathrm{measure}\ j|\mathrm{truth\ is}\ i)$, thus it can be used to optimize the measurement results.

\begin{figure*}[h!]
\centering
\includegraphics[scale = 0.6]{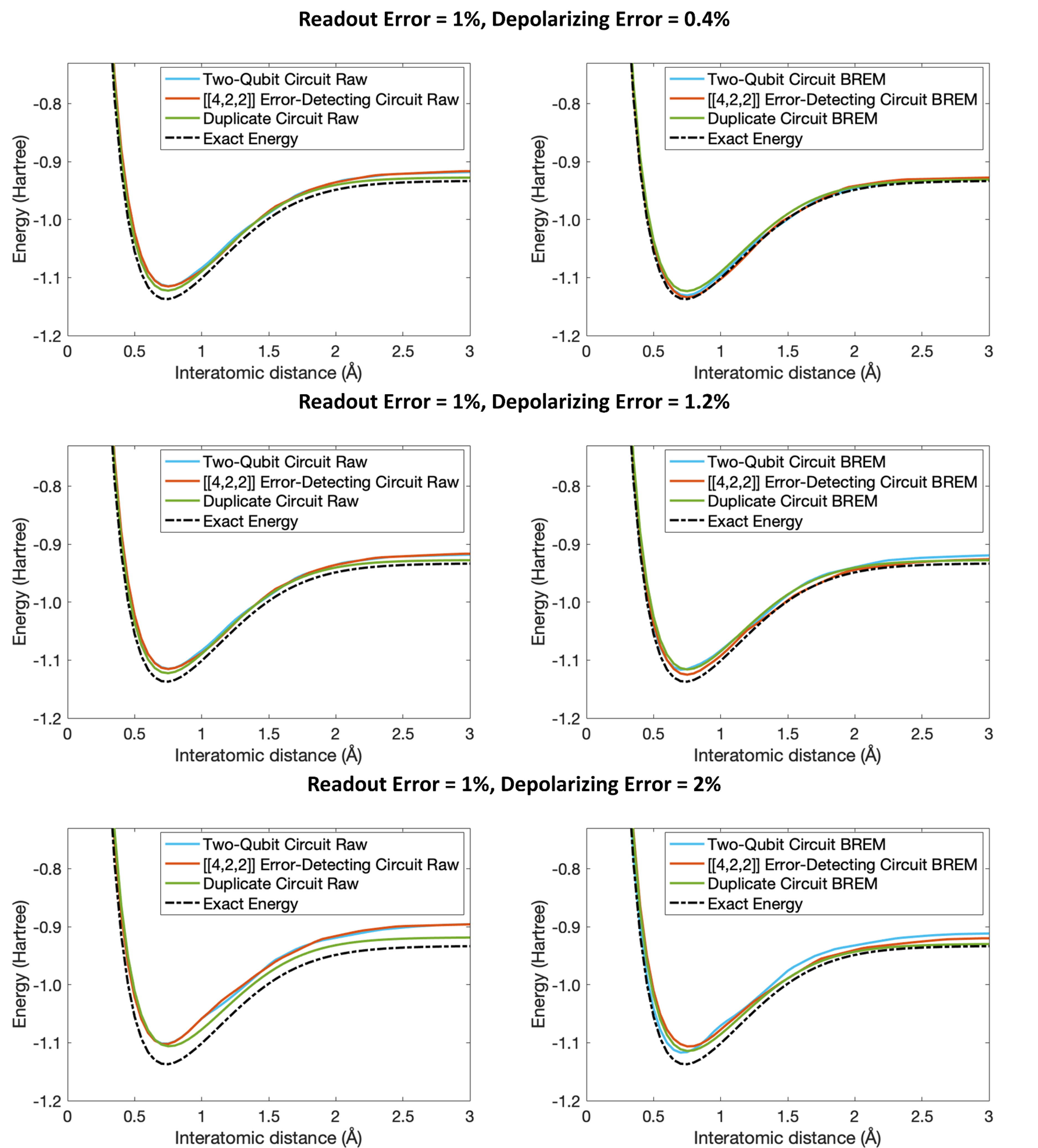}
\caption[VQE Results on Mixed Noisy Simulator with Fixed read-out error Rate 0.01\%]{The comparison of VQE results of three circuits on the mixed noisy simulator with the combination of three depolarizing error rates (0.4\%, 1.2\% and 2\%) and fixed read-out error rate 1\%.}
\label{fig:alldeprd1}
\end{figure*}

Furthermore, we observe that the raw result curve for duplicate circuit show the best performance compare to that of the other two circuits in a large error rate (2\%). The positive effects of this approach outweigh the negative effects of this approach, that is, the noise introduced by 30 gates does not affects to calculate an accurate expectation value by amplifying the largest eigenvalue exponentially, which verifies our assumption in Fig.~\ref{fig:energy_errors}; the optimized curve for the QEC method shows improved accuracy compared to the corresponding raw curve, since after the optimization, we discarded the states with non-zero ancilla value and renormalized the probability of remaining states. When the noise level is high and the BREM optimization does not work, many incorrect states can be discarded during these two procedures, the probability of correct states move closer to the corresponding exact value, which can improve the fidelity of optimized curves in the QEC method. 

\begin{figure*}[h!]
\centering
\includegraphics[scale = 0.6]{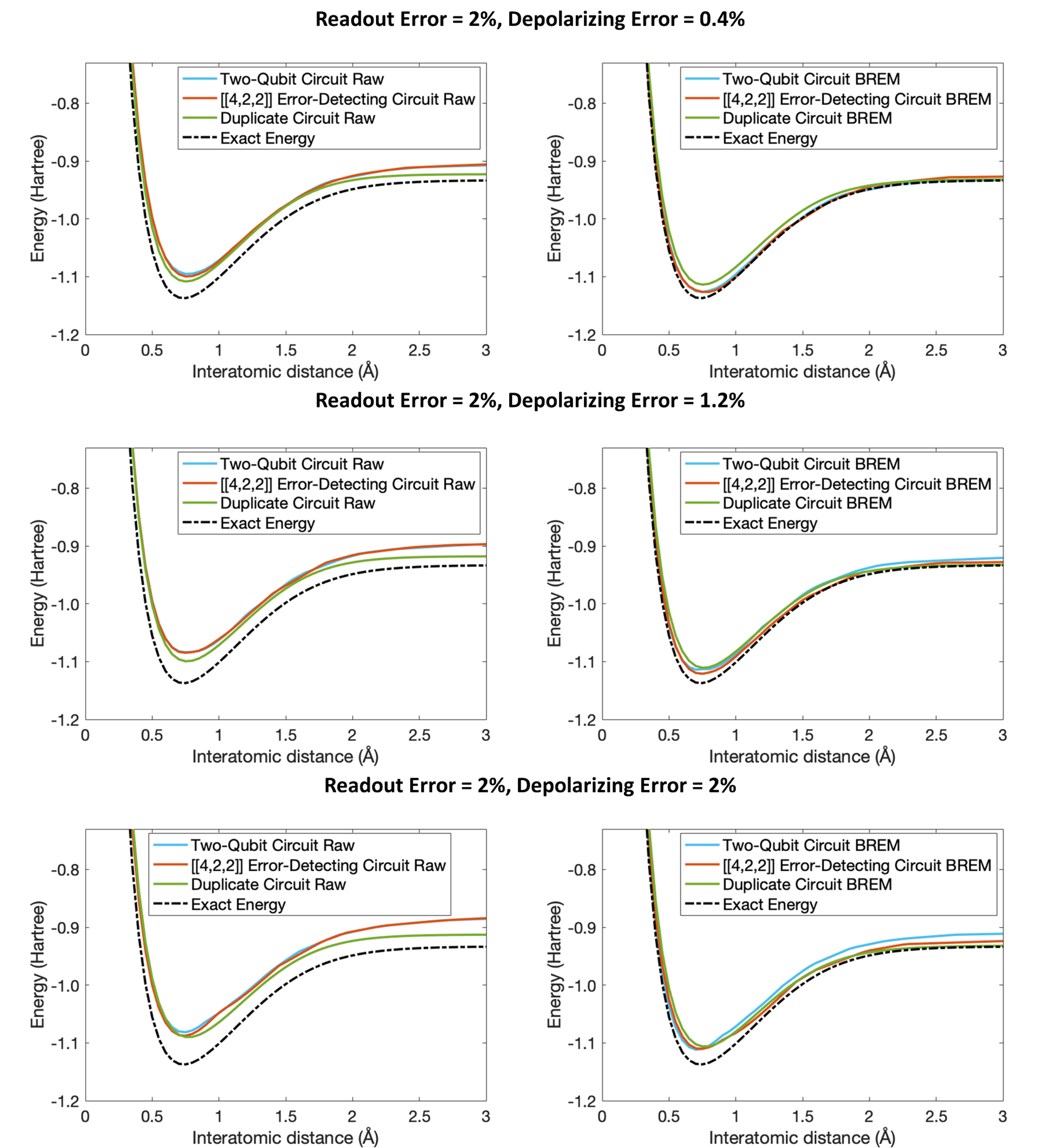}
\caption[VQE Results on Mixed Noisy Simulator with Fixed read-out error Rate 0.02\%]{The comparison of VQE results of three circuits on the mixed noisy simulator with the combination of three depolarizing error rates (0.4\%, 1.2\% and 2\%) and fixed read-out error rate 2\%.}
\label{fig:alldeprd2}
\end{figure*}

\begin{figure*}[h!]
\centering
\includegraphics[scale = 0.6]{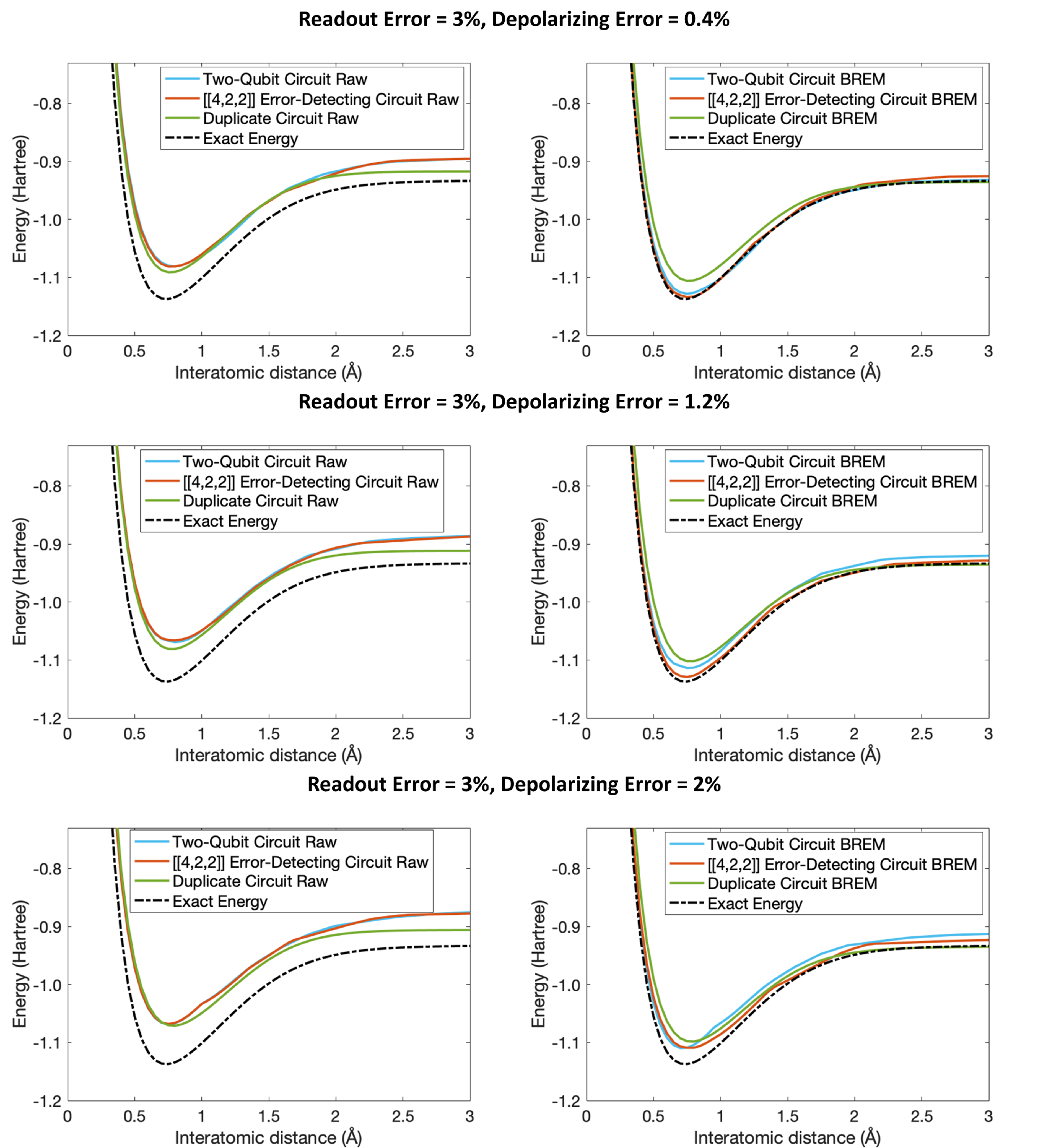}
\caption[VQE Results on Mixed Noisy Simulator with Fixed read-out error Rate 0.03\%]{The comparison of VQE results of three circuits on the mixed noisy simulator with the combination of three depolarizing error rates (0.4\%, 1.2\% and 2\%) and fixed read-out error rate 3\%.}
\label{fig:alldeprd3}
\end{figure*}

\subsubsection{Relative Performance of Error Mitigation Schemes}

To further quantify the performance of the QEC method and duplicate circuits, we increase depolarisation error rate from 0 to 12\% and compute the ground state energy at a selected inter-atomic distance of $\SI{0.75}{\angstrom}$. In these simulations, we do not include any measurement or read-out errors. Figure~\ref{fig:energy_errors} shows the error in ground-state energy of H$_2$ molecule as a function of quantum gate error rate for each circuit. The results show that the duplicate circuit and the QEC method both outperform the two-qubit circuit when the gate error rate is increased from 0 to 12\%, which is significantly larger than the actual gate error rate on $ibmq\_toronto$ (around 8\% for CNOT error). Furthermore, duplicate circuit out-performs the QEC method for the whole noise range. The inset in Fig.~\ref{fig:energy_errors} shows that the two-qubit circuit implementation performs better than the duplicate circuit and QEC method only when error rate is increased above 32\%, which is unrealistic for practical devices.

Quantum devices generally exhibit much large error rate for two-qubit gates compared to single qubit gates, therefore, we also simulated duplicate circuit with only two-qubit gate noise (ideal single qubit gates). Our results in Fig.~\ref{fig:energy_errors} indicate that such scenario results in only a small improvement in the accuracy. Finally, we simulated duplicate circuit with ideal (zero noise) $B$ gates, whereas the actual VQE circuit consists of noisy single and two-qubit gates. In this scenario, we find that the accuracy of energy calculation is drastically improved. Based on these results, we conclude that the performance of duplicate circuit can be drastically enhanced by mapping $B$ gates on least noise qubits available on a quantum processor.   

\subsubsection{Depolarizing Noise and read-out errors}

For the third set of simulations, we added both depolarizing gate noise and read-out errors to a noisy simulator to test the performance of the BREM optimization on three circuits shown in Fig.~\ref{fig:all_circs}. The computed results are summarised in Fig.~\ref{fig:alldeprd1}, Fig.~\ref{fig:alldeprd2} and Fig.~\ref{fig:alldeprd3}, where each figure compares the VQE results from the three circuits on a simulator with the combination of a fixed read-out error rate and three depolarizing error rates. Overall, we find that the implementation of BREM scheme improves the fidelity of all three circuits. However, interestingly, the improvement from BREM is much larger for two-qubit circuit and QEC method when compared to the duplicate circuit. This is clearly evident from the results at 3\% read-out error results, where the duplicate circuit wins based on the raw data, but is no longer the best option after the BREM implementation. Another finding from our results is that the implementation of BREM technique is less effective in the presence of both depolarising and read-out noise when compared to both type of noise individually present. We also conclude that the hardware noise is quite different from simple depolarisation noise and read-out errors. Importantly, the performance of duplicate circuit is extremely well when implemented on IBM device, compared to noisy simulations, which is more relevant for practical applications. 

\section{Conclusions and Future Directions}

In this work, we have provided a comprehensive analysis of the performance of the three error mitigation techniques in the context of proof-of-concept H$_2$ molecule ground-state energy calculations. Our results provide key new insights into the working of the studied error mitigation schemes based on both quantum hardware experiments and noisy simulation environment with varying strengths and types of noise models. When the three circuits are executed on $ibmq\_toronto$, the duplicate circuit gives significantly better performance when compared to the two-qubit implementation and the QEC method. We also find that the BREM scheme is very efficient in mitigating the impact of read-out errors. When we implemented multiple mappings simultaneously on quantum devices, our results indicate that strong cross-talk impacts both duplicate circuit and the QEC experiments. To gain further insights of the comparative working of the three error mitigation schemes under various noise configurations, we performed simulations with varying degrees of depolarising and read-out errors. Overall the simulated results support the findings from the hardware experiments that the duplicate circuit approach provides the best fidelity by mitigating the impact of depolarisation noise. 

We also identify a few directions for future research. First, one of the differences between $ibmq\_toronto$ quantum processor and a noisy simulator is that actual device has complex type of errors not fully captured by simple depolarising error models. Future investigations can focus on the performance of three circuits on complex error models including cross-talk and spatially/temporally inhomogeneous and correlated noise. Secondly, based on the experimental results of the duplicate circuit in this paper, we found that the positive effects in calculating the expectation value of the subsystem outweigh the negative effects of noise brought by bringing more quantum gates. Future work can focus to expand this study by working in two directions: one is to test this method in a quantum system with more subsystems by designing a complicated cascaded $B$ gate; another one is to focus on designing the $B$ gate efficiently, to introduce less noise to the system. Although the duplicate circuit in our work shows excellent performance when compared against both hardware noise and simulated error models, its benchmarking for larger molecular systems such as LiH or BeH$_2$ will be an important topic of future research.

In summary, our results provide key new insights into the working of the duplicate circuit, the QEC method and BREM technique for noise mitigation in quantum chemistry simulations which is anticipated as one of the killer applications of quantum computing.  

\begin{acknowledgments}

We acknowledge the use of IBM Quantum services for this work. The views expressed are those of the authors, and do not reflect the official policy or position of IBM or the IBM Quantum team.

\end{acknowledgments}

\bibliography{reference}



\end{document}